\documentclass[a4paper,aps,twocolumn,nofootinbib]{revtex4}
\RequirePackage[colorlinks,hyperindex]{hyperref}
\RequirePackage[english]{babel}
\RequirePackage[latin1]{inputenc}
\RequirePackage[T1]{fontenc}
\RequirePackage{mathrsfs}
\RequirePackage{amsmath}
\RequirePackage{amssymb}
\RequirePackage{amsbsy}
\RequirePackage{color}
\RequirePackage{bm}
\hypersetup{colorlinks=true,breaklinks=true,urlcolor=blue,linkcolor=red}
\begin{document}
%%%%%%%%%%%%%%%%%%%%%%%%%%%%%%%%%%%%%%%%%%%%%%%%%%%%%%%%%%%%%%%%%%%%%%%%%%%%%%%%%%%%%%%%%%%%%%%%%%%
\title{\bf{Polar Form of Spinor Fields from Regular to Singular: the Flag-Dipoles}}
\author{Luca Fabbri$^{\hbar}$ and Rodolfo Jos\'{e} Bueno Rogerio$^{\nabla}$}
\affiliation{$^{\hbar}$DIME Sez. Metodi e Modelli Matematici, Universit\`{a} di
Genova, via all'Opera Pia 15, 16145 Genova, ITALY\\
$^{\nabla}$Institute of Physics and Chemistry, Federal 
University of Itajub\'{a}, Itajub\'{a}, Minas Gerais, 37500-903, BRAZIL}
\date{\today}
%%%%%%%%%%%%%%%%%%%%%%%%%%%%%%%%%%%%%%%%%%%%%%%%%%%%%%%%%%%%%%%%%%%%%%%%%%%%%%%%%%%%%%%%%%%%%%%%%%%
\begin{abstract}
In this paper, we perform the polar analysis of the spinorial fields, starting from the regular cases and up to the singular cases: we will give for the first time the polar form of the spinorial field equations for the singular cases constituted by the flag-dipole spinor fields. Comments on the role of further spinor sub-classes containing Majorana and Weyl spinors will be sketched.
\end{abstract}
%%%%%%%%%%%%%%%%%%%%%%%%%%%%%%%%%%%%%%%%%%%%%%%%%%%%%%%%%%%%%%%%%%%%%%%%%%%%%%%%%%%%%%%%%%%%%%%%%%%
\maketitle
%%%%%%%%%%%%%%%%%%%%%%%%%%%%%%%%%%%%%%%%%%%%%%%%%%%%%%%%%%%%%%%%%%%%%%%%%%%%%%%%%%%%%%%%%%%%%%%%%%%
\section{Introduction}
One of the most important building blocks in modern physics is the spinor field. Originally constituted by the Dirac field alone, it has been later recognized that Dirac spinors are only one part of a more varied population of spinorial fields. The Dirac spinor can in fact be defined as the spinor whose scalar and pseudo-scalar bi-linear quantities are not both identically equal to zero. However, this leaves the door open for an altogether different type of spinors having both scalar and pseudo-scalar bi-linear quantities vanishing identically, called singular \cite{L,Cavalcanti:2014wia}.

Singular spinor fields, or flag-dipole spinor fields, may be unusual but they still have many important things to tell \cite{HoffdaSilva:2017waf,daSilva:2012wp,Cavalcanti:2020obq,Ablamowicz:2014rpa,daRocha:2008we, Meert:2018qzk,Rogerio:2019xcu,Rogerio:2020ewe}. In fact, these spinors can be further split into sub-classes, obtained when the axial-vector bi-linear quantity, that is the spin, is zero, and in this case they are called flagpole spinors, or when the antisymmetric tensor bi-linear quantity, that is the momentum, is zero, and in this case they are called dipole spinors. The flagpole and dipole spinors are important because they respectively contain Majorana and Weyl spinors. In particular, Majorana spinors have a considerable relevance due to their double-helicity structure as it was discussed in \cite{Ahluwalia:2004sz,Ahluwalia:2004ab,Ahluwalia:2016rwl,Ahluwalia:2016jwz,daRocha:2007pz,HoffdaSilva:2009is,Cavalcanti:2014uta,Villalobos:2015xca,daRocha:2011yr}.

This full classification of spinor fields points toward the fact that it may be less and less wise to focus on just one class of spinors forgetting about the rest. From a mathematical point of view, one would have to investigate all such classes in order to decide which ones can be taken as basis for the physical description of the nature we know.

One mathematical tool that has been recently investigated is the so-called polar form of spinor fields. The polar form of spinor fields is merely the way we have to write spinor fields in a form in which each component is expressed as a module times a complex phase while still maintaining manifest covariance. After such a transformation, the spinor field and all the associated quantities are written in terms of real tensors only. What is most important is the fact that in this polar form, spinor fields exhibit the true degrees of freedom separated from all other components, with the possibility to introduce new tensors describing the background interactions, and to write the polar form of the field equations too \cite{Rodrigues:2005yz,Fabbri:2016msm,Fabbri:2018crr}.

The polar form of the spinor field equations in the regular case has been found in \cite{Fabbri:2016laz,Fabbri:2017pwp}, but for the singular case we only had partial analyses \cite{Fabbri:2017xyk,Fabbri:2019vut}. In the present paper, however, we intend to straighten this situation by providing the full polar decomposition of the spinor field equations in the singular case, the flag-dipole case.

We will also discuss sub-cases, highlighting how flagpole and dipole spinorial classes are extinguished by the Majorana and Weyl spinors respectively. We shall also comment, however, that these types of spinors are much more singular than previously thought since they do not possess any true degree of freedom after all.
%%%%%%%%%%%%%%%%%%%%%%%%%%%%%%%%%%%%%%%%%%%%%%%%%%%%%%%%%%%%%%%%%%%%%%%%%%%%%%%%%%%%%%%%%%%%%%%%%%%
%%%%%%%%%%%%%%%%%%%%%%%%%%%%%%%%%%%%%%%%%%%%%%%%%%%%%%%%%%%%%%%%%%%%%%%%%%%%%%%%%%%%%%%%%%%%%%%%%%%
\section{General Spinors}
As a start, we recall that $\boldsymbol{\gamma}^{a}$ are the Clifford matrices, from which $\left[\boldsymbol{\gamma}_{a},\!\boldsymbol{\gamma}_{b}\right]\!=\!4\boldsymbol{\sigma}_{ab}$ and $2i\boldsymbol{\sigma}_{ab}\!=\!\varepsilon_{abcd}\boldsymbol{\pi}\boldsymbol{\sigma}^{cd}$ are the definitions of the $\boldsymbol{\sigma}_{ab}$ and the $\boldsymbol{\pi}$ matrices. This last is as a norm indicated by a gamma with an index five, but in the space-time this index has no meaning, and therefore we will use a notation with no index. The former instead are complex matrices that verify the Lorentz algebra, and as a consequence they are the generators of the complex Lorentz group. Because parameters are point-dependent, this group will be a gauge group. Objects $\psi$ transforming in terms of the transformations of this group are called spinor fields. We can define an adjoint spinor field according to $\overline{\psi}\!=\!\psi^{\dagger}\boldsymbol{\gamma}^{0}$ because in such a way we have that
\begin{eqnarray}
&2\overline{\psi}\boldsymbol{\sigma}^{ab}\boldsymbol{\pi}\psi\!=\!\Sigma^{ab}\\
&2i\overline{\psi}\boldsymbol{\sigma}^{ab}\psi\!=\!M^{ab}\\
&\overline{\psi}\boldsymbol{\gamma}^{a}\boldsymbol{\pi}\psi\!=\!S^{a}\\
&\overline{\psi}\boldsymbol{\gamma}^{a}\psi\!=\!U^{a}\\
&i\overline{\psi}\boldsymbol{\pi}\psi\!=\!\Theta\\
&\overline{\psi}\psi\!=\!\Phi
\end{eqnarray}
are all objects that transform in terms of the real Lorentz transformations. For these bi-linear quantities, we have
\begin{eqnarray}
&\Sigma^{ab}\!=\!-\frac{1}{2}\varepsilon^{abij}M_{ij}\\
&M^{ab}\!=\!\frac{1}{2}\varepsilon^{abij}\Sigma_{ij}
\end{eqnarray}
and
\begin{eqnarray}
&M_{ab}\Phi\!-\!\Sigma_{ab}\Theta\!=\!U^{j}S^{k}\varepsilon_{jkab}\label{A1}\\
&M_{ab}\Theta\!+\!\Sigma_{ab}\Phi\!=\!U_{[a}S_{b]}\label{A2}
\end{eqnarray}
together with
\begin{eqnarray}
&\frac{1}{2}M_{ab}M^{ab}\!=\!-\frac{1}{2}\Sigma_{ab}\Sigma^{ab}\!=\!\Phi^{2}\!-\!\Theta^{2}
\label{norm2}\\
&\frac{1}{2}M_{ab}\Sigma^{ab}\!=\!-2\Theta\Phi
\label{orthogonal2}\\
&U_{a}U^{a}\!=\!-S_{a}S^{a}\!=\!\Theta^{2}\!+\!\Phi^{2}\label{norm1}\\
&U_{a}S^{a}\!=\!0\label{orthogonal1}
\end{eqnarray}
which are called Fierz re-arrangement identities.

There is a point we have to make now, and it is about the adjoint defined by the $\overline{\psi}\!=\!\psi^{\dagger}\boldsymbol{\gamma}^{0}$ above. This adjoint, or dual, can be proven to be uniquely defined, up to a re-naming of all the bi-linear spinor quantities, if we assume that the dualization be done universally. However, one might drop this assumption and require that dualization be performed in a momentum-dependent way \cite{Cavalcanti:2020obq}.

From the metric, we define the symmetric connection as usual with $\Lambda^{\sigma}_{\alpha\nu}$ from which, with the tetrads, we define the spin connection $\Omega^{a}_{b\pi}\!=\!\xi^{\nu}_{b}\xi^{a}_{\sigma}(\Lambda^{\sigma}_{\nu\pi}\!-\!\xi^{\sigma}_{i}\partial_{\pi}\xi_{\nu}^{i})$ from which, with the gauge potential, we define the spinor connection
\begin{eqnarray}
&\boldsymbol{\Omega}_{\mu}
=\frac{1}{2}\Omega^{ab}_{\phantom{ab}\mu}\boldsymbol{\sigma}_{ab}
\!+\!iqA_{\mu}\boldsymbol{\mathbb{I}}\label{spinorialconnection}
\end{eqnarray}
in general. With it we can define
\begin{eqnarray}
&\boldsymbol{\nabla}_{\mu}\psi\!=\!\partial_{\mu}\psi
\!+\!\boldsymbol{\Omega}_{\mu}\psi\label{spincovder}
\end{eqnarray}
as the spinorial covariant derivative.

We notice that there is no torsion in such a derivative, which is therefore not the most general possible. Nevertheless, we can recover full generality by adding torsion as an axial-vector potential in the dynamics.

For this we will take the Dirac equation
\begin{eqnarray}
&i\boldsymbol{\gamma}^{\mu}\boldsymbol{\nabla}_{\mu}\psi
\!-\!XW_{\sigma}\boldsymbol{\gamma}^{\sigma}\boldsymbol{\pi}\psi\!-\!m\psi\!=\!0
\label{D}
\end{eqnarray}
where $W_{\nu}$ is the torsion axial-vector. Multiplying these on the left by all Clifford matrices $\mathbb{I}$, $\boldsymbol{\pi}$, $\boldsymbol{\gamma}^{i}$, $\boldsymbol{\gamma}^{i}\boldsymbol{\pi}$, $\boldsymbol{\sigma}^{ij}$ and the adjoint spinor field, then splitting real and imaginary parts, we obtain the following expressions
\begin{eqnarray}
&\frac{i}{2}(\overline{\psi}\boldsymbol{\gamma}^{\mu}\boldsymbol{\nabla}_{\mu}\psi
\!-\!\boldsymbol{\nabla}_{\mu}\overline{\psi}\boldsymbol{\gamma}^{\mu}\psi)
\!-\!XW_{\sigma}S^{\sigma}\!-\!m\Phi\!=\!0\\
&\nabla_{\mu}U^{\mu}\!=\!0
\end{eqnarray}
\begin{eqnarray}
&\frac{i}{2}(\overline{\psi}\boldsymbol{\gamma}^{\mu}\boldsymbol{\pi}\boldsymbol{\nabla}_{\mu}\psi
\!-\!\boldsymbol{\nabla}_{\mu}\overline{\psi}\boldsymbol{\gamma}^{\mu}\boldsymbol{\pi}\psi)
\!-\!XW_{\sigma}U^{\sigma}\!=\!0\\
&\nabla_{\mu}S^{\mu}\!-\!2m\Theta\!=\!0
\end{eqnarray}
\begin{eqnarray}
\nonumber
&\frac{i}{2}(\overline{\psi}\boldsymbol{\nabla}^{\alpha}\psi
\!-\!\boldsymbol{\nabla}^{\alpha}\overline{\psi}\psi)
\!-\!\frac{1}{2}\nabla_{\mu}M^{\mu\alpha}-\\
&-\frac{1}{2}XW_{\sigma}M_{\mu\nu}\varepsilon^{\mu\nu\sigma\alpha}\!-\!mU^{\alpha}\!=\!0
\label{vr}\\
\nonumber
&\nabla_{\alpha}\Phi
\!-\!2(\overline{\psi}\boldsymbol{\sigma}_{\mu\alpha}\!\boldsymbol{\nabla}^{\mu}\psi
\!-\!\!\boldsymbol{\nabla}^{\mu}\overline{\psi}\boldsymbol{\sigma}_{\mu\alpha}\psi)+\\
&+2X\Theta W_{\alpha}\!=\!0\label{vi}
\end{eqnarray}
\begin{eqnarray}
\nonumber
&\nabla_{\nu}\Theta\!-\!
2i(\overline{\psi}\boldsymbol{\sigma}_{\mu\nu}\boldsymbol{\pi}\boldsymbol{\nabla}^{\mu}\psi\!-\!
\boldsymbol{\nabla}^{\mu}\overline{\psi}\boldsymbol{\sigma}_{\mu\nu}\boldsymbol{\pi}\psi)-\\
&-2X\Phi W_{\nu}\!+\!2mS_{\nu}\!=\!0\label{ar}\\
\nonumber
&(\boldsymbol{\nabla}_{\alpha}\overline{\psi}\boldsymbol{\pi}\psi
\!-\!\overline{\psi}\boldsymbol{\pi}\boldsymbol{\nabla}_{\alpha}\psi)
\!-\!\frac{1}{2}\nabla^{\mu}M^{\rho\sigma}\varepsilon_{\rho\sigma\mu\alpha}+\\
&+2XW^{\mu}M_{\mu\alpha}\!=\!0\label{ai}
\end{eqnarray}
\begin{eqnarray}
\nonumber
&\nabla^{\mu}S^{\rho}\varepsilon_{\mu\rho\alpha\nu}
\!+\!i(\overline{\psi}\boldsymbol{\gamma}_{[\alpha}\!\boldsymbol{\nabla}_{\nu]}\psi
\!-\!\!\boldsymbol{\nabla}_{[\nu}\overline{\psi}\boldsymbol{\gamma}_{\alpha]}\psi)+\\
&+2XW_{[\alpha}S_{\nu]}\!=\!0\\
\nonumber
&\nabla^{[\alpha}U^{\nu]}\!+\!i\varepsilon^{\alpha\nu\mu\rho}
(\overline{\psi}\boldsymbol{\gamma}_{\rho}\boldsymbol{\pi}\boldsymbol{\nabla}_{\mu}\psi\!-\!\!
\boldsymbol{\nabla}_{\mu}\overline{\psi}\boldsymbol{\gamma}_{\rho}\boldsymbol{\pi}\psi)-\\
&-2XW_{\sigma}U_{\rho}\varepsilon^{\alpha\nu\sigma\rho}\!-\!2mM^{\alpha\nu}\!=\!0
\end{eqnarray}
which are known as Gordon-Madelung decompositions.

Again, it must be noted that not all spinors are defined dynamically in terms of the Dirac equations, as dual-helicity spinors might obey Klein-Gordon equations \cite{Ahluwalia:2016jwz}.

This development is general. Now is time to perform the classification of spinor fields, and for that we will closely follow the Lounesto classification, based on the bi-linear spinor fields \cite{L,Cavalcanti:2014wia,HoffdaSilva:2017waf,daSilva:2012wp,Cavalcanti:2020obq,Ablamowicz:2014rpa,daRocha:2008we, Meert:2018qzk,Rogerio:2019xcu,Rogerio:2020ewe}. Since all the bi-linear spinors are tensors, such a classification, based on vanishing these tensors, is manifestly generally covariant. As a start, we split the cases obtained either when at least one between $\Theta$ or $\Phi$ is not identically zero and giving rise to the \emph{regular} spinors or when $\Theta\!=\!\Phi\!\equiv\!0$ giving rise the \emph{singular} spinors in general. Regular spinors are what contains the Dirac spinors (general cases are class-I, while $\Theta\!=\!0$ defines the class-II and $\Phi\!=\!0$ defines the class-III). Singular spinors can be further split in three classes, according to whether $M^{ab}$ and $S^{a}$ are not identically zero giving the \emph{flag-dipole} spinors (class-IV), or by having $S^{a}\!\equiv\!0$ giving the \emph{flagpole} spinors (class-V, also containing Majorana spinors) while $M^{ab}\!\equiv\!0$ gives the \emph{dipole} spinors (class-VI, also containing Weyl spinors). Notice that $U^{a}\!=\!0$ results into $\psi\!=\!0$ and so the Lounesto classification is proven to be exhaustive.

We will split these cases in the following sub-sections.
%%%%%%%%%%%%%%%%%%%%%%%%%%%%%%%%%%%%%%%%%%%%%%%%%%%%%%%%%%%%%%%%%%%%%%%%%%%%%%%%%%%%%%%%%%%%%%%%%%%
\subsection{Regular Spinors}
Regular spinors are those defined in general when at least one of the two scalars is non-zero.

In this case (\ref{norm1}) tells that $U^{a}$ is time-like, so that we can always perform up to three boosts to bring its spatial components to vanish. Then, it is always possible to use the rotation around the first and second axes to bring the space part of $S^{a}$ aligned with the third axis. And finally, it is always possible to employ the last rotation to bring the spinor into the form
\begin{eqnarray}
&\!\!\psi\!=\!\phi e^{-\frac{i}{2}\beta\boldsymbol{\pi}}
\boldsymbol{S}\left(\!\begin{tabular}{c}
$1$\\
$0$\\
$1$\\
$0$
\end{tabular}\!\right)
\label{regular}
\end{eqnarray}
where $\boldsymbol{S}$ is a general spinor transformation and this is what is called polar form of regular spinor fields. In polar form regular spinor fields are such that
\begin{eqnarray}
&\Sigma^{ab}\!=\!2\phi^{2}(\cos{\beta}u^{[a}s^{b]}\!-\!\sin{\beta}u_{j}s_{k}\varepsilon^{jkab})\\
&M^{ab}\!=\!2\phi^{2}(\cos{\beta}u_{j}s_{k}\varepsilon^{jkab}\!+\!\sin{\beta}u^{[a}s^{b]})
\end{eqnarray}
showing that the antisymmetric tensors are written in terms of the vectors
\begin{eqnarray}
&S^{a}\!=\!2\phi^{2}s^{a}\\
&U^{a}\!=\!2\phi^{2}u^{a}
\end{eqnarray}
and the scalars
\begin{eqnarray}
&\Theta\!=\!2\phi^{2}\sin{\beta}\\
&\Phi\!=\!2\phi^{2}\cos{\beta}
\end{eqnarray}
with constraints $u_{a}u^{a}\!=\!-s_{a}s^{a}\!=\!1$ and $u_{a}s^{a}\!=\!0$ showing that the velocity vector $u^{a}$ and spin-axial vector $s^{a}$ are fixed and thus the scalar $\phi$ and the pseudo-scalar $\beta$ are the only degrees of freedom of the system, being called module and Yvon-Takabayashi angle respectively \cite{Fabbri:2016msm}.

With regular spinor fields in polar form it is not difficult to see that $\boldsymbol{S}$ is generally given with the structure
\begin{eqnarray}
&\boldsymbol{S}\partial_{\mu}\boldsymbol{S}^{-1}\!=\!i\partial_{\mu}\sigma\mathbb{I}
\!+\!\frac{1}{2}\partial_{\mu}\theta_{ij}\boldsymbol{\sigma}^{ij}\label{parameters}
\end{eqnarray}
where $\sigma$ is a generic complex phase and $\theta_{ij}\!=\!-\theta_{ji}$ are the six parameters of the Lorentz group, so that we define
\begin{eqnarray}
&\partial_{\mu}\theta_{ij}\!-\!\Omega_{ij\mu}\!\equiv\!R_{ij\mu}\label{R}\\
&\partial_{\mu}\sigma\!-\!qA_{\mu}\!\equiv\!P_{\mu}\label{P}
\end{eqnarray}
which can be proven to be real tensors and therefore they are called \emph{tensorial connection} and \emph{gauge-invariant vector momentum}. With them
\begin{eqnarray}
&\!\boldsymbol{\nabla}_{\mu}\psi\!=\!(\nabla_{\mu}\ln{\phi}\mathbb{I}
\!-\!\frac{i}{2}\nabla_{\mu}\beta\boldsymbol{\pi}
\!-\!iP_{\mu}\mathbb{I}\!-\!\frac{1}{2}R_{ij\mu}\boldsymbol{\sigma}^{ij})\psi
\label{decspinder}
\end{eqnarray}
from which
\begin{eqnarray}
&\nabla_{\mu}s_{i}\!=\!R_{ji\mu}s^{j}\label{ds}\\
&\nabla_{\mu}u_{i}\!=\!R_{ji\mu}u^{j}\label{du}
\end{eqnarray}
are valid as general geometric identities \cite{Fabbri:2018crr}.

Finally, by substituting the polar form of spinorial covariant derivative of regular spinor fields into the Gordon decompositions and setting
\begin{eqnarray}
&\frac{1}{2}\varepsilon_{\mu\alpha\nu\iota}R^{\alpha\nu\iota}\!=\!B_{\mu}\\
&R_{\mu a}^{\phantom{\mu a}a}\!=\!R_{\mu}
\end{eqnarray}
one can isolate the pair of independent field equations
\begin{eqnarray}
&B_{\mu}\!-\!2P^{\iota}u_{[\iota}s_{\mu]}
\!-\!2XW_{\mu}\!+\!\nabla_{\mu}\beta\!+\!2s_{\mu}m\cos{\beta}\!=\!0\label{f1}\\
&R_{\mu}\!-\!2P^{\rho}u^{\nu}s^{\alpha}\varepsilon_{\mu\rho\nu\alpha}
\!+\!2s_{\mu}m\sin{\beta}\!+\!\nabla_{\mu}\ln{\phi^{2}}\!=\!0\label{f2}
\end{eqnarray}
specifying all derivatives of both degrees of freedom, and as it is possible to see they imply (\ref{D}), so that (\ref{f1}, \ref{f2}) are equivalent to the Dirac spinorial field equations \cite{Fabbri:2016laz}.

All this for regular spinors is now very well established so we proceed to do the same analysis for singular spinors.
%%%%%%%%%%%%%%%%%%%%%%%%%%%%%%%%%%%%%%%%%%%%%%%%%%%%%%%%%%%%%%%%%%%%%%%%%%%%%%%%%%%%%%%%%%%%%%%%%%%
\subsection{Singular Spinors: Flag-Dipoles}
Singular spinors are defined by $\Theta\!=\!\Phi\!=\!0$ identically and their polar decomposition was first studied in \cite{Fabbri:2017xyk} and \cite{Fabbri:2019vut}, so here we plan to perform a full analysis.

In this case (\ref{norm2}, \ref{orthogonal2}) tell that if $M_{ab}$ is written in terms of $M_{0K}\!=\!E_{K}$ and $M_{IJ}\!=\!\varepsilon_{IJK}B^{K}$ then they are such that $E^{2}\!=\!B^{2}$ and $\vec{E}\!\cdot\!\vec{B}\!=\!0$ and employing the same reasoning used above we have that we can always boost and rotate them so to bring $\vec{B}$ and $\vec{E}$ aligned respectively with the first and second axis. Then, the spinor field is
\begin{eqnarray}
&\!\!\psi\!=\!\frac{1}{\sqrt{2}}(\mathbb{I}\cos{\frac{\alpha}{2}}
\!-\!\boldsymbol{\pi}\sin{\frac{\alpha}{2}})\boldsymbol{S}\left(\!\begin{tabular}{c}
$1$\\
$0$\\
$0$\\
$1$
\end{tabular}\!\right)
\label{singular}
\end{eqnarray}
with $\boldsymbol{S}$ a general spin transformation and this is called polar form of singular spinor fields. In polar form singular spinors verify the relationships
\begin{eqnarray}
&S^{a}\!=\!-\sin{\alpha}U^{a}\label{m}
\end{eqnarray}
with $U_{a}U^{a}\!=\!0$ as well as $M_{ab}M^{ab}\!=\!\varepsilon^{abij}M_{ab}M_{ij}\!=\!0$ and $M_{ik}U^{i}\!=\!\varepsilon_{ikab}M^{ab}U^{i}\!=\!0$ and therefore showing that the pseudo-scalar $\alpha$ is the only degree of freedom we have.

With singular spinor fields in polar form and again
\begin{eqnarray}
&\boldsymbol{S}\partial_{\mu}\boldsymbol{S}^{-1}\!=\!i\partial_{\mu}\sigma\mathbb{I}
\!+\!\frac{1}{2}\partial_{\mu}\theta_{ij}\boldsymbol{\sigma}^{ij}
\end{eqnarray}
we can still define
\begin{eqnarray}
&\partial_{\mu}\theta_{ij}\!-\!\Omega_{ij\mu}\!\equiv\!R_{ij\mu}\\
&\partial_{\mu}\sigma\!-\!qA_{\mu}\!\equiv\!P_{\mu}
\end{eqnarray}
as above. With them, for singular spinor fields we have
\begin{eqnarray}
\nonumber
&\boldsymbol{\nabla}_{\mu}\psi\!=\![-\frac{1}{2}(\mathbb{I}\tan{\alpha}
\!+\!\boldsymbol{\pi}\sec{\alpha})\nabla_{\mu}\alpha-\\
&-iP_{\mu}\mathbb{I}\!-\!\frac{1}{2}R_{ij\mu}\boldsymbol{\sigma}^{ij}]\psi
\end{eqnarray}
from which
\begin{eqnarray}
&\!\!\nabla_{\mu}U_{i}\!=\!R_{ji\mu}U^{j}\\
&\!\!\!\!\nabla_{\mu}M^{ab}\!=\!-M^{ab}\tan{\alpha}\nabla_{\mu}\alpha
\!-\!R^{a}_{\phantom{a}k\mu}M^{kb}\!+\!R^{b}_{\phantom{b}k\mu}M^{ka}
\end{eqnarray}
are valid as general geometric identities.

Finally, by substituting the polar form of spinorial covariant derivative of singular spinor fields into the Gordon decompositions, then plugging the bi-linear spinorial quantities, and diagonalizing the results, we get
\begin{eqnarray}
&\!\!(\varepsilon^{\mu\rho\sigma\nu}\nabla_{\mu}\alpha\sec{\alpha}
\!-\!2P^{[\rho}g^{\sigma]\nu})M_{\rho\sigma}\!=\!0
\label{1}\\
&\!\!\!\!M_{\rho\sigma}(g^{\nu[\rho}\nabla^{\sigma]}\alpha\sec{\alpha}
\!-\!2P_{\mu}\varepsilon^{\mu\rho\sigma\nu})\!+\!4m\sin{\alpha}U^{\nu}\!=\!0
\label{2}
\end{eqnarray}
\begin{eqnarray}
\nonumber
&\!\![(2XW\!-\!B)^{\sigma}\varepsilon_{\sigma\mu\rho\nu}\!+\!R_{[\mu}g_{\rho]\nu}+\\
&+g_{\nu[\mu}\nabla_{\rho]}\alpha\tan{\alpha}]M_{\eta\zeta}\varepsilon^{\mu\rho\eta\zeta}\!=\!0
\label{3}\\
\nonumber
&\!\!\!\![(2XW\!-\!B)^{\sigma}\varepsilon_{\sigma\mu\rho\nu}\!+\!R_{[\mu}g_{\rho]\nu}+\\
&+g_{\nu[\mu}\nabla_{\rho]}\alpha\tan{\alpha}]M^{\mu\rho}\!+\!4mU_{\nu}\!=\!0
\label{4}
\end{eqnarray}
specifying all derivatives of the degree of freedom, and as it turns out they are equivalent to the initial Dirac spinorial field equations, as we are now going to demonstrate.

To see that, it will be easier to go in the frame where $\boldsymbol{S}\!=\!\mathbb{I}$ in which the bi-linear spinor quantities reduce to
\begin{eqnarray}
&M^{02}\!=\!\cos{\alpha}\ \ \ \ M^{23}\!=\!\cos{\alpha}\label{w}\\
&U^{0}\!=\!1\ \ \ \ U^{3}\!=\!-1
\end{eqnarray}
in a form that is important for the field equations.

In fact, the above equations in this frame become
\begin{eqnarray}
&P_{0}=P_{3}\\
&R_{0}=R_{3}\\
&(B\!-\!2XW)_{0}\!=\!(B\!-\!2XW)_{3}\\
&(2XW\!-\!B)_{1}\!-\!R_{2}+2P_{1}\sin{\alpha}\!+\!2m\cos{\alpha}\!=\!0\\
&(2XW\!-\!B)_{2}\!+\!R_{1}+2P_{2}\sin{\alpha}\!=\!0\\
&\nabla_{0}\alpha=\nabla_{3}\alpha\\
&\nabla_{1}\alpha+2P_{2}\cos{\alpha}=0\\
&\nabla_{2}\alpha\!-\!2P_{1}\cos{\alpha}\!+\!2m\sin{\alpha}\!=\!0
\end{eqnarray}
showing that we do specify all derivatives of the degree of freedom, and in this form it is also possible to see that they imply the Dirac spinorial field equations in polar form in that very frame. Because a spinor field equation valid in a frame is also valid in every frame then they imply the Dirac spinorial field equations in polar form, and thus they imply the Dirac spinorial field equations in general. Thus we can conclude that they are totally equivalent to the initial Dirac spinorial field equations.

It would still be interesting to know why in this case, albeit the equivalence is obtained, nevertheless the polar form of the field equations displays such a redundancy in the number of apparently independent field equations.

Solutions to these equations have first been found in references \cite{Vignolo:2011qt,daRocha:2013qhu} without employing methods involving the polar form, and it is our belief that with the polar form the quest for further solutions might be simplified.

There are two especially singular cases we also have to split, and we are going to do it in the following.

\subsubsection{Zero-Spin Flag-Dipoles: the Flagpoles (Majorana)}
A first special sub-class of flag-dipoles is the one for which $\alpha\!=\!0$ and in this case the spinor field reduces to
\begin{eqnarray}
&\!\!\psi\!=\!\frac{1}{\sqrt{2}}\boldsymbol{S}\left(\!\begin{tabular}{c}
$1$\\
$0$\\
$0$\\
$1$
\end{tabular}\!\right)
\end{eqnarray}
with
\begin{eqnarray}
&S^{a}\!=\!0
\end{eqnarray}
having $U_{a}U^{a}\!=\!0$ as well as $M_{ab}M^{ab}\!=\!\varepsilon^{abij}M_{ab}M_{ij}\!=\!0$ and $M_{ik}U^{i}\!=\!\varepsilon_{ikab}M^{ab}U^{i}\!=\!0$ so that $\boldsymbol{\gamma}^{2}\psi^{\ast}\!=\!\eta\psi$ with $\eta$ a constant unitary phase added for generality and which is the well known case of the Majorana spinorial field.

Conversely, condition $\boldsymbol{\gamma}^{2}\psi^{\ast}\!=\!\eta\psi$ with $\eta\!=\!\mp i$ gives
\begin{eqnarray}
&\!\!\!\!(\mathbb{I}\cos{\frac{\alpha}{2}}
\!+\!\boldsymbol{\pi}\sin{\frac{\alpha}{2}})\left(\!\begin{tabular}{c}
$1$\\
$0$\\
$0$\\
$1$
\end{tabular}\!\right)\!=\!\pm(\mathbb{I}\cos{\frac{\alpha}{2}}
\!-\!\boldsymbol{\pi}\sin{\frac{\alpha}{2}})\left(\!\begin{tabular}{c}
$1$\\
$0$\\
$0$\\
$1$
\end{tabular}\!\right)
\end{eqnarray}
then $\alpha\!=\!\pi\!+\!2\pi n$ or $\alpha\!=\!2\pi n$ so that because of (\ref{m}) we necessarily have the zero-spin flag-dipole spinorial field.

As a consequence, zero-spin flag-dipoles, also known as flagpoles, form a class that is exhausted by the Majorana spinorial field (up to a local Lorentz transformation).

It is however important to notice that there remains no degree of freedom in this very specific situation.

\subsubsection{Zero-Momentum Flag-Dipoles: the Dipoles (Weyl)}
Another special sub-class of flag-dipoles is the one for which $\alpha\!=\!\pm\pi/2$ so that the spinor field reduces to
\begin{eqnarray}
&\!\!\psi\!=\!\frac{1}{2}(\mathbb{I}\mp\boldsymbol{\pi})\boldsymbol{S}\left(\!\begin{tabular}{c}
$1$\\
$0$\\
$0$\\
$1$
\end{tabular}\!\right)
\end{eqnarray}
that is 
\begin{eqnarray}
&\!\!\psi\!=\!\boldsymbol{S}\left(\!\begin{tabular}{c}
$1$\\
$0$\\
$0$\\
$0$
\end{tabular}\!\right)\ \ \ \ \mathrm{or} \ \ \ \ \psi\!=\!\boldsymbol{S}\left(\!\begin{tabular}{c}
$0$\\
$0$\\
$0$\\
$1$
\end{tabular}\!\right)
\end{eqnarray}
with
\begin{eqnarray}
&S^{a}\!=\!\mp U^{a}
\end{eqnarray}
that is
\begin{eqnarray}
&S^{a}\!=\!-U^{a}\ \ \ \ \mathrm{or} \ \ \ \ S^{a}\!=\!U^{a}
\end{eqnarray}
for the left-handed and right-handed chiral part respectively, with $U_{a}U^{a}\!=\!0$ as well as $M_{ab}\!=\!0$ so that the spinor has become an eigen-state of the chiral projectors and this is the well known case of the pair of left-handed and right-handed Weyl chiral spinorial fields.

Conversely, to select eigen-spinors of the chiral projectors it is necessary that 
\begin{eqnarray}
&\frac{1}{\sqrt{2}}(\mathbb{I}\cos{\frac{\alpha}{2}}
\!-\!\boldsymbol{\pi}\sin{\frac{\alpha}{2}})
\!=\!\frac{1}{2}(\mathbb{I}\!\mp\!\boldsymbol{\pi})
\end{eqnarray}
then we have $\cos{\frac{\alpha}{2}}\!=\!1/\sqrt{2}$ and $\sin{\frac{\alpha}{2}}\!=\!\pm1/\sqrt{2}$ and finally $\alpha\!=\!\pm\pi/2\!+\!2\pi n$ so that because of (\ref{w}) we must have the zero-momentum flag-dipole spinor field in general cases.

As a consequence, zero-momentum flag-dipoles, also known as dipoles, form a class that is exhausted by the Weyl spinorial field (up to local Lorentz transformations).

It is however important to notice that there remains no degree of freedom in this very specific situation too.
%%%%%%%%%%%%%%%%%%%%%%%%%%%%%%%%%%%%%%%%%%%%%%%%%%%%%%%%%%%%%%%%%%%%%%%%%%%%%%%%%%%%%%%%%%%%%%%%%%%
%%%%%%%%%%%%%%%%%%%%%%%%%%%%%%%%%%%%%%%%%%%%%%%%%%%%%%%%%%%%%%%%%%%%%%%%%%%%%%%%%%%%%%%%%%%%%%%%%%%
\section{Physical Considerations}
\subsection{Regular Case}
Having presented the general geometry of spinor fields, it is our next goal to take into account physical considerations involving them, and we will split the two cases of regular versus singular for easier comparison, starting with the regular case. In this case, the spinor is characterized by two degrees of freedom, Yvon-Takabayashi angle and module, whose interpretation is straightforward.

The module is in fact the same module we have in non-relativistic limit and it measures the density of the material distribution. The Yvon-Takabayashi angle is a measure of the phase difference between the two chiral parts of the spinor, and therefore it encodes the information about the internal dynamics of the spinor itself.

This can be seen by going into the rest-frame of the spinor and noticing that even there it would still be impossible to have the non-relativistic limit if $\beta$ does not identically vanish. So relativistic effects taking place in rest-frame can only be due to internal motions. As such the Yvon-Takabayashi angle is intrinsically relativistic.

Computing the energy of the spinor would result into
\begin{eqnarray}
\nonumber
&\!\!\frac{i}{2}(\overline{\psi}\boldsymbol{\gamma}^{\nu}\boldsymbol{\nabla}_{\mu}\psi
\!-\!\boldsymbol{\nabla}_{\mu}\overline{\psi}\boldsymbol{\gamma}^{\nu}\psi)
\!=\!2\phi^{2}[s^{\nu}\nabla_{\mu}\beta/2+\\
&+u^{\nu}P_{\mu}\!+\!\frac{1}{4}\varepsilon^{\rho\alpha\sigma\nu}s_{\rho}R_{\alpha\sigma\mu}]
\end{eqnarray}
in which we see that apart from the usual term $u^{\nu}P_{\mu}$ there are two terms involving the contribution of the spin, one of which being proportional to the covariant gradient of the Yvon-Takabayashi angle giving the contribution of the internal dynamics. Notice that all is weighted by the module, which for square-integrable solutions reduces the energy density to zero at the infinity of the space.

It is interesting at this point to speculate about some possible special values that the Yvon-Takabayashi angle can assume. In case the Yvon-Takabayashi angle is constant the energy density loses all internal contributions.

This is to be expected. In fact $\beta\!=\!n\pi$ corresponds to the Lounesto class-II, which is made up by spinors whose two chiral parts are equal, therefore without relative motions. Instead $\beta\!=\!\pi/2\!+\!n\pi$ gives the Lounesto class-III, made up by spinors that can be thought as some possible superposition of spinors in the pure-particle state $\beta\!=\!0$ plus spinors in the pure-antiparticle state $\beta\!=\!\pi$ \cite{Recami:1995iy,Salesi:1995vy}.

Thus, Lounesto classes II and III constitute some sort of singular sub-classes within the regular classes.
%%%%%%%%%%%%%%%%%%%%%%%%%%%%%%%%%%%%%%%%%%%%%%%%%%%%%%%%%%%%%%%%%%%%%%%%%%%%%%%%%%%%%%%%%%%%%%%%%%%
\subsection{Singular Case --- Diverging Quantities}
The singular case is not so straightforward, however.

It amounts to a single degree of freedom because the module is boosted away with a boost along the direction of the motion. Thus, we should expect that the algebraic conserved quantities fail to behave as densities.

In fact, for example, the velocity does not go to zero at infinity, and in fact it even remains constant, creating the problem that once integrated over the volume it would give rise to divergent conserved quantities. This problem might be overcome by requiring that such a quantity be unobservable, that is that such a quantity be not related to any physical source. If this has to be the case, then, we are in the situation in which we cannot define a coupling to electrodynamics, or at least not as usual. Similar considerations apply also to the spin as source of torsion.

Different is the case of the energy since it is differentially related to the field. Computing it gives
\begin{eqnarray}
\nonumber
&\frac{i}{2}(\overline{\psi}\boldsymbol{\gamma}^{\nu}\boldsymbol{\nabla}_{\mu}\psi
\!-\!\boldsymbol{\nabla}_{\mu}\overline{\psi}\boldsymbol{\gamma}^{\nu}\psi)\!=\!U^{\nu}P_{\mu}+\\
&+\frac{1}{4}\varepsilon^{\rho\alpha\sigma\nu}S_{\rho}R_{\alpha\sigma\mu}
\end{eqnarray}
in which all derivatives of $\alpha$ have disappeared. The lack of a module weighting the energy density might indicate that even for the energy there may be convergence problems after volume integration. Its trace
\begin{eqnarray}
&\frac{i}{2}(\overline{\psi}\boldsymbol{\gamma}^{\mu}\boldsymbol{\nabla}_{\mu}\psi
\!-\!\boldsymbol{\nabla}_{\mu}\overline{\psi}\boldsymbol{\gamma}^{\mu}\psi)\!=\!
-\frac{1}{2}\sin{\alpha}U^{\rho}B_{\rho}
\end{eqnarray}
is the only non-trivial scalar that is related to a conserved quantity which can be computed in this model.

Notice that in the Lounesto classes V and VI, flagpole and dipole spinors reduce to have zero energy trace, in the former case due to condition $\sin{\alpha}\!=\!0$ and in the latter case because Weyl spinors are necessarily massless.
%%%%%%%%%%%%%%%%%%%%%%%%%%%%%%%%%%%%%%%%%%%%%%%%%%%%%%%%%%%%%%%%%%%%%%%%%%%%%%%%%%%%%%%%%%%%%%%%%%%
%%%%%%%%%%%%%%%%%%%%%%%%%%%%%%%%%%%%%%%%%%%%%%%%%%%%%%%%%%%%%%%%%%%%%%%%%%%%%%%%%%%%%%%%%%%%%%%%%%%
\section{Comparing Models}
So far we have presented the way to perform the spinor field polar decomposition, splitting regular and singular cases. The most dramatic difference involving these cases is that while regular cases are characterized by a module that may go to zero at infinity fast enough to make the integrals over the volume of all conserved quantities converge, singular cases do not have such a module on which to rely for integrals of conserved quantities to converge.

As such it may appear that singular spinors might be used to build a physically meaningful theory only if the theory itself displays some dynamical restrictions.

Another interesting question is whether it is possible to find some mapping that would convert a regular spinor into a singular spinor in general circumstances \cite{R1,R2}.

We are going to leave this question to a following work.
%%%%%%%%%%%%%%%%%%%%%%%%%%%%%%%%%%%%%%%%%%%%%%%%%%%%%%%%%%%%%%%%%%%%%%%%%%%%%%%%%%%%%%%%%%%%%%%%%%%
%%%%%%%%%%%%%%%%%%%%%%%%%%%%%%%%%%%%%%%%%%%%%%%%%%%%%%%%%%%%%%%%%%%%%%%%%%%%%%%%%%%%%%%%%%%%%%%%%%%
\section{Conclusion}
In this paper, we have considered the Lounesto classification of spinors, separating regular spinor fields from the singular spinor fields characterized by $\Theta\!=\!\Phi\!=\!0$ called flag-dipole spinors. And these are further split into sub-classes, one in which also $S^{a}\!=\!0$ called flagpole and containing Majorana spinors and one in which also $M^{ab}\!=\!0$ called dipole and containing two Weyl chiral spinors.

Of all classes, we provided the polar decomposition of the field equations. The polar form of all the field equations for regular spinors was already given in \cite{Fabbri:2016laz} and here we have given the polar form of the field equations for the singular flag-dipole. We have also discussed flagpole and dipole cases, showing how in polar form it is easy to see that the former extinguishes the class of Majorana spinors while the latter extinguishes the class of Weyl spinors, and proving that neither has true degrees of freedom, as it was already anticipated in \cite{Fabbri:2017xyk}.

Opportunities for further development may be found in exploiting the polar form of field equations (\ref{1}, \ref{2}, \ref{3}, \ref{4}) to find more general solutions. Other works may involve finding a way to map regular into singular spinors.

The last of these two problems is already being investigated for an up-coming work.
%%%%%%%%%%%%%%%%%%%%%%%%%%%%%%%%%%%%%%%%%%%%%%%%%%%%%%%%%%%%%%%%%%%%%%%%%%%%%%%%%%%%%%%%%%%%%%%%%%%
%%%%%%%%%%%%%%%%%%%%%%%%%%%%%%%%%%%%%%%%%%%%%%%%%%%%%%%%%%%%%%%%%%%%%%%%%%%%%%%%%%%%%%%%%%%%%%%%%%%

%%%%%%%%%%%%%%%%%%%%%%%%%%%%%%%%%%%%%%%%%%%%%%%%%%%%%%%%%%%%%%%%%%%%%%%%%%%%%%%%%%%%%%%%%%%%%%%%%%%
\end{document}